\documentclass[12pt]{article}

\pdfoutput=1

\usepackage{cite}

\usepackage{amssymb,latexsym}

\usepackage{epstopdf}

\usepackage{graphics,epsfig}

\usepackage{color}
\usepackage{xcolor}

\usepackage{amsmath,amsfonts}

\usepackage{mathrsfs}

\DeclareMathAlphabet{\mathpzc}{OT1}{pzc}{m}{it}

\usepackage{tikz}
\usetikzlibrary{arrows,shapes}
\usetikzlibrary{trees}
\usetikzlibrary{matrix,arrows} 				
\usetikzlibrary{positioning}				
\usetikzlibrary{calc,through}				
\usetikzlibrary{decorations.pathreplacing}  
\usepackage{pgffor}							

\usetikzlibrary{decorations.pathmorphing}	
\usetikzlibrary{decorations.markings}
\tikzset{
    vector/.style={decorate, decoration={snake}, draw},
	provector/.style={decorate, decoration={snake,amplitude=2.5pt}, draw},
	antivector/.style={decorate, decoration={snake,amplitude=-2.5pt}, draw},
    fermion/.style={draw=black, postaction={decorate},
        decoration={markings,mark=at position .55 with {\arrow[draw=black]{>}}}},
    fermionbar/.style={draw=black, postaction={decorate},
        decoration={markings,mark=at position .55 with {\arrow[draw=black]{<}}}},
    fermionnoarrow/.style={draw=black},
    gluon/.style={decorate, draw=black,
        decoration={coil,amplitude=4pt, segment length=5pt}},
    scalar/.style={dashed,draw=black, postaction={decorate},
        decoration={markings,mark=at position .55 with {\arrow[draw=black]{>}}}},
    scalarbar/.style={dashed,draw=black, postaction={decorate},
        decoration={markings,mark=at position .55 with {\arrow[draw=black]{<}}}},
    scalarnoarrow/.style={dashed,draw=black},
    electron/.style={draw=black, postaction={decorate},
        decoration={markings,mark=at position .55 with {\arrow[draw=black]{>}}}},
	bigvector/.style={decorate, decoration={snake,amplitude=4pt}, draw},
}

\tikzstyle{block} = [draw, rectangle, 
    minimum height=3em, minimum width=6em]

\let\a=\alpha \let\b=\beta \let\g=\gamma \let\d=\delta \let\e=\epsilon
\let\z=\zeta  \let\th=\theta  \let\k=\kappa
\let\l=\lambda \let\m=\mu \let\n=\nu \let\x=\xi \let\p=\pi 
\let\s=\sigma   \let\f=\phi  
 
      \let\G=\Gamma \let\D=\Delta \let\Th=\Theta 
\let\X=\Xi  \let\S=\Sigma  \let\Y=\Psi
 
\let\la=\label  
  
\def\nn{\nonumber} \def\bd{\begin{document}} \def\ed{\end{document}}
\def\ds{\documentstyle} \let\fr=\frac \let\bl=\bigl \let\br=\bigr
\let\Br=\Bigr \let\Bl=\Bigl
\let\bm=\bibitem
\let\na=\nabla
\def\tU{{\widetilde U}}
\let\pa=\partial \let\ov=\overline
\def\ie{{\it i.e.\ }}
\newcommand{\be}{\begin{equation}}
\newcommand{\ee}{\end{equation}}
\def\ba{\begin{array}}
\def\ea{\end{array}}
\def\ft#1#2{{\textstyle{{\scriptstyle #1}\over {\scriptstyle #2}}}}
\def\fft#1#2{{#1 \over #2}}
\def\F#1#2{{ F_{#1}^{(#2)} }}
\def\cF#1#2{{ {\cal F}_{#1}^{(#2)} }}

\def\R{{\bf R}}
\def\sst#1{{\scriptscriptstyle #1}}
\def\oneone{\rlap 1\mkern4mu{\rm l}}
\def\e7{E_{7(+7)}}
\def\td{\tilde}
\def\wtd{\widetilde}
\def\im{{\rm i}}
\def\bog{Bogomol'nyi\ }
\newcommand{\ho}[1]{$\, ^{#1}$}
\newcommand{\hoch}[1]{$\, ^{#1}$}
\newcommand{\bea}{\begin{eqnarray}}
\newcommand{\eea}{\end{eqnarray}}
\newcommand{\ra}{\rightarrow}
\newcommand{\lra}{\longrightarrow}
\newcommand{\Lra}{\Leftrightarrow}
\newcommand{\ap}{\alpha^\prime}
\newcommand{\bp}{\tilde \beta^\prime}
\newcommand{\cB}{{\cal B}}
\newcommand{\cO}{{\cal O}}
\newcommand{\vecx}{\vec{x}}
\newcommand{\vecy}{\vec{y}}
\newcommand{\vecp}{\vec{p}}
\newcommand{\vecq}{\vec{q}}
\newcommand{\tr}{{\rm tr} }
\newcommand{\Tr}{{\rm Tr} }
\newcommand{\NP}{Nucl. Phys. }

\newcommand{\cL}{{\cal L}}
\newcommand{\cA}{{\cal A}}
\newcommand{\cT}{{\cal T}}
\newcommand{\cR}{{\cal R}}
\newcommand{\cD}{{\cal D}}
\newcommand{\cH}{{\cal H}}

\def\Cb{\bar{C}}

\def\sst#1{{\scriptscriptstyle #1}}
\def\0{{\sst{(0)}}}
\def\1{{\sst{(1)}}}
\def\2{{\sst{(2)}}}
\def\3{{\sst{(3)}}}
\def\4{{\sst{(4)}}}
\def\5{{\sst{(5)}}}
\def\6{{\sst{(6)}}}
\def\7{{\sst{(7)}}}
\def\8{{\sst{(8)}}}
\def\9{{\sst{(9)}}}
\def\p{{\sst{(p)}}}
\def\q{{\sst{(q)}}}
\def\ve{\varepsilon}
\def\vf{\varphi}
\def\F{\Phi}
\def\wg{\wedge}

\def\thb{\bar{\theta}}
\def\Thb{\bar{\Theta}}
\def\barp{\bar{p}}
\def\barq{\bar{q}}
\def\barc{\bar{c}}
\def\bard{\bar{d}}
\def\e{\epsilon}

\def \bi{\bibitem}
\def \la {\label}

\def \l {\lambda}
\def\foot{\footnote}
\def \tl  {{\tilde \l}}
\def \sql {{\sqrt \l}}
\def \adss {$AdS_5 \times S^5$\ }
\newcommand{\rf}[1]{(\ref{#1})}
\def \ov {\over}

\def\th{\theta}
\def\Th{\Theta}
\def\vth{\vartheta}
\def\btheta{{\bar\theta}}
\def\ttheta{{{\tilde\theta}}}
\def\bttheta{{{\bar\ttheta}}}
\def\vth{\vartheta}

\def\ra{\rightarrow}
\def\N{\nabla}
\def\F{{\cal F}}
\def\uM{\underline{M}}
\def\uA{\underline{A}}
\def\uN{\underline{N}}
\def\uP{\underline{P}}
\def\ua{\underline{a}}
\def\ub{\underline{b}}
\def\uc{\underline{c}}
\def\ud{\underline{d}}
\def\ue{\underline{e}}
\def\uf{\underline{f}}
\def\ui{\underline{i}}
\def\uj{\underline{j}}
\def\uk{\underline{k}}
\def\ul{\underline{l}}
\def\ual{\underline{\alpha}}
\def\ube{\underline{\beta}}
\def\um{\underline{m}}
\def\un{\underline{n}}
\def\up{\underline{p}}
\def\uq{\underline{q}}
\def\ur{\underline{r}}
\def\us{\underline{s}}
\def\umu{\underline{\mu}}
\def\unu{\underline{\nu}}
\def\ula{\underline{\l}}
\def\uka{\underline{\k}}
\def\usi{\underline{\s}}
\def\urh{\underline{\r}}
\def\cc{\circ}
\def\eqv{\equiv}

\def\ni{\noindent}

\def\Ep{E^{{}^{(+)}}}
\def\Em{E^{{}^{(-)}}}

\def\Mp{M^{{}^{(+)}}}
\def\Mm{M^{{}^{(-)}}}

\def \ha{{1\ov 2}}

\def\r{\rho}

\def\Y{{\rm Y}}
\def\X{{\rm X}}
\def\tY{\tilde{\rm Y}}
\def\tX{\tilde{\rm X}}
\def\dY{\dot{\rm Y}}
\def\dX{\dot{\rm X}}

\def \J {\mathcal{J}}
\def \del {\partial}

\def\dF{\dot{F}}
\def\dG{\dot{G}}
\def\df{\dot{f}}
\def \E {{\cal E}}
\def \S {{\cal S}}
\def \J {{\cal J}}

\def\ms{\mathcal{S}}
\def\mj{\mathcal{J}}
\def\soj{\fr{\ms}{\mj}}
\def \R {{\bf R}}
\def \om {\omega}
\def \bE {\bar E}
\def \x {{\cal X}}

\def \bi{\bibitem}
\def \la {\label}

\def \l {\lambda}
\def\foot{\footnote}
\def \tl  {{\tilde \l}}
\def \sql {{\sqrt \l}}
\def \adss {$AdS_5 \times S^5$\ }
\def \ov {\over}

\def \varpi {{\rm w}}

\def\thb{\bar{\theta}}
\def\Thb{\bar{\Theta}}
\def\mb{\bar{\m}}
\def\ab{\bar{\a}}
\def\zb{\bar{z}}
\def\psib{\bar{\psi}}
\def\barp{\bar{p}}
\def\barq{\bar{q}}
\def\barc{\bar{c}}
\def\bard{\bar{d}}
\def\e{\epsilon}
\def\wb{\bar{w}}
\def\lb{\bar{\l}}
\def\Jb{\bar{J}}
\def\Nb{\bar{N}}
\def\Zb{\bar{Z}}
\def\pab{\bar{\pa}}

\def\At{\tilde{A}}
\def\Bt{\tilde{B}}
\def\Ct{\tilde{C}}
\def\Dt{\tilde{D}}
\def\Et{\tilde{E}}
\def\Ft{\tilde{F}}
\def\Gt{\tilde{G}}
\def\Ht{\tilde{H}}
\def\Kt{\tilde{K}}
\def\Mt{\tilde{M}}
\def\Nt{\tilde{N}}
\def\Rt{\tilde{R}}
\def\at{\tilde{a}}
\def\bt{\tilde{b}}
\def\ct{\tilde{c}}
\def\dt{\tilde{d}}
\def\et{\tilde{e}}
\def\ft{\tilde{f}}
\def \ztt{\tilde{\z}}
\def \zetat{\tilde{\zeta}}
\def\htil{\tilde{h}}
\def\gt{\tilde{g}}
\def\nt{\tilde{n}}
\def\mut{\tilde{\mu}}
\def\nut{\tilde{\nu}}
\def\pht{\tilde{\f}}
\def\Phit{\tilde{\Phi}}
\def\vft{\tilde{\vf}}

\def\rht{\tilde{\rho}}

\def\asth{\hat{*}}
\def\phh{\hat{\phi}}

\def\bA{{\bf A}}

\def\ola{\overleftarrow}
\def\ora{\overrightarrow}
\def\alt{\tilde{\a}}

\def\eh{\hat{e}}
\def\eph{\hat{\e}}
\def\ph{\hat{p}}
\def\alh{\hat{\a}}
\def\beh{\hat{\b}}
\def\gah{\hat{\g}}
\def\Fh{\hat{F}}
\def\muh{\hat{\m}}
\def\nuh{\hat{\n}}
\def\thh{\hat{\th}}
\def\rhh{\hat{\r}}
\def\dh{\hat{d}}
\def\ih{\hat{i}}
\def\jh{\hat{j}}
\def\hh{\hat{h}}
\def\nh{\hat{n}}
\def\gh{\hat{g}}
\def\kh{\hat{k}}
\def\deh{\hat{\d}}
\def\wh{\hat{w}}
\def\lah{\hat{\l}}
\def\Ah{\hat{A}}
\def\Kh{\hat{K}}
\def\Nh{\hat{N}}
\def\Rh{\hat{R}}
\def\Ch{\hat{C}}
\def\Omh{\hat{\Omega}}

\def\xh{\hat{x}}

\def\ps{\rlap{\, /}\;\,p }
\def\ks{\rlap{\, /}\;\,k }

\def\gym{g_{YM}}

\def\adot{\dot{a}}
\def\bdot{\dot{b}}
\def\bpa{\bar{\pa}}

\def\pr{\prime}
\def\ssk{\medskip}
\def\clb{\color{blue}}
\def\clr{\color{red}}
\def\clg{\color{green}}
\def\clp{\color{purple}}
\def\clc{\color{cyan}}
\def\clm{\color{magenta}}
\def\cly{\color{yellow}}

\def\bfA{{\bf A}}
\def\bfB{{\bf B}}
\def\bfK{{\bf K}}
\def\bfU{{\bf U}}
\def\bfX{{\bf X}}
\def\bfY{{\bf Y}}
\def\bfZ{{\bf Z}}
\def\bfg{{\bf g}}
\def\bfn{{\bf n}}

\def\bsk{\bigskip}
\def\ssk{\medskip}

\def\Ec{{\cal E}}

\begin{document}

\overfullrule=0pt
\parskip=2pt
\parindent=12pt
\headheight=0in \headsep=0in \topmargin=0in
\oddsidemargin=0in

\vspace{ -3cm}
\thispagestyle{empty}

 \vspace{0.1cm}

\setcounter{equation}{0}
\setcounter{footnote}{0}
\setcounter{section}{0}

\begin{center}

{\Large\bf  Boundary dynamics in gravitational theories}

\vskip 0.8cm

%
%
I. Y. Park
\\
\vspace{0.3cm}
%
%
\vspace{0.3cm}
{\it Department of Applied Mathematics,
Philander Smith College 
                               \\
Little Rock, AR 72223, USA \\
inyongpark05@gmail.com
}

 \vspace{.5cm}

\end{center}

 \vspace{0.1cm}

\begin{abstract}

We present a foliation-focused critical review of the boundary conditions and dynamics of 4D gravitational theories. A general coordinate transformation introduces a new foliation and changes the hypersurface on which a natural boundary condition is imposed; in this sense gauge transformations must be viewed as changing the boundary conditions. The issue of a gauge invariant boundary condition is nontrivial and has been extensively studied in the literature. We turn around the difficulty in obtaining such a boundary condition (and subtleties observed in the main body) and take it as one of the indications of an enlarged Hilbert space so as to include the states satisfying different boundary conditions. Through the systematical reduction procedure we obtain, up to some peculiarities, the explicit form of the reduced Lagrangian that describes the dynamics of the physical states. 
We examine the new insights offered by the 3D Lagrangian on BMS-type symmetry and black hole information. In particular we confirm that the boundary dynamics is an indispensable part of the system information.

\end{abstract}
\newpage





\section{Introduction}

One of the key lessons learned from the recent developments in theoretical physics is that the degrees of freedom on the boundary are as important as those of the bulk.\footnote{The discussions of the boundary conditions and dynamics go back to the pre-AdS/CFT era \cite{Regge:1974zd,Benguria:1976in,Witten:1988hf}.} This is true not only for the anti-de Sitter-based dualities such as the AdS/CFT correspondence, but also for far more general cases, implying the potential existence of a large class of holographic dualities. The importance of the boundary degrees of freedom naturally raises the issues germane to the boundary conditions and dynamics (earlier works and reviews can be found, e.g., in \cite{Barvinsky:1987dg,Espositobook,Avramidi:1997sh,Papadimitriou:2004ap,vanNieuwenhuizen:2005kg,Moss:2013vh,Jacobson:2013yqa}) whose significances have not been fully recognized. In the present work we review the boundary conditions and dynamics, highlight their salient features, and clarify various issues. The boundary theory Lagrangian that describes the dynamics of the physical degrees of freedom will also be obtained through a systematical reduction procedure. With the explicit reduced Lagrangian obtained, many aspects of the holographic description of the 4D theory become more transparent.

Although the AdS/CFT correspondence now has a well-established dictionary between the two dual theories involved, it was lacking the {\em explicit} dualization procedure that converts one to the other. Several such examples of the dualization procedures were studied in \cite{Park:2001bm,Park:2008sg,Hatefi:2012bp,Park:2017wiw} (see also \cite{Niarchos:2015moa,Grignani:2016bpq,Maxfield:2016vpw}). In particular, in \cite{Park:2008sg,Hatefi:2012bp,Park:2017wiw}, the gauge theory - whose explicit form was worked out in  \cite{Hatefi:2012bp} - was understood as the generalized Goldstone degrees of freedom of the gravity theory. It was also realized through a series of works that what is responsible for the holographic property is the large amount of gauge symmetry: in the ADM formalism \cite{Arnowitt:1962hi}, the Hamiltonian and momentum constraints can be explicitly solved for relatively simple backgrounds with the solution implying that the physical sector of the theory is associated with the 3D boundary hypersurface \cite{Park:2014tia}\cite{Park:2015ybl}. The reduction has been further generalized to include more general backgrounds \cite{Park:2017wiw}.
To some extent the present work is in a vein similar to that of \cite{Hatefi:2012bp}: the boundary theory is explicitly computed. The 3D metric, being the ``moduli field"  \cite{Hatefi:2012bp} \cite{Park:2017wiw}, represents the generalized Goldstone degrees of freedom.

One of the focuses of the present work is the foliation-focused implications of the gauge transformations for the boundary conditions and dynamics. As a matter of fact, there exists extensive literature on boundary conditions in a gauge theory, see, e.g, \cite{Barvinsky:1987dg,Espositobook,Avramidi:1997sh,Papadimitriou:2004ap,vanNieuwenhuizen:2005kg,Moss:2013vh,Jacobson:2013yqa} and references therein. The issue of the gauge invariant boundary terms turns out to be very nontrivial although it is possible to come up with such boundary terms and carry out more rigorous analysis in certain cases, see, e.g., \cite{Papadimitriou:2004ap}\cite{Moss:2013vh}. Moreover, as will be discussed in the main body the presence of large gauge transformations (LGTs) will pull one out of the physical content of the original boundary condition although the forms of the (gauge invariant) boundary terms will be maintained. This status of matter leads us to propose that one should consider an enlarged Hilbert space including the sectors coming from different boundary conditions.

Among other things what distinguishes the present work from the existing literature is its focus on the relevance of foliation of the geometry and complementarity of the active and passive transformations.\footnote{{The definitions of the active and passive transformations will be given in the body.}} 
A coordinate transformation introduces a new foliation, which is directly relevant for the boundary conditions. 
Due to the new slicing associated with the new coordinates, the {\em natural} boundary hypersurface in the new coordinates is not, in general, the one obtained by transforming the boundary hypersurface in the original coordinates. Thus, the corresponding Gibbons-Hawking-York (GHY) term is different from the original GHY-term in, say, the foliation content although its form remains the same.
In this sense, a gauge transformation should be viewed as changing the boundary condition {even if it is a non-large, i.e., ordinary, one.} (Our view on this matter will be articulated in section 2.)
One of the implications of the boundary dynamics and gauge transformations is the relevance of the boundary conditions different from a Dirichlet boundary condition (a related idea can be found in \cite{Papadimitriou:2005ii}), in particular the relevance of Neumann-type boundary conditions.
It is then imperative to come up with a new setup in which all those different boundary conditions are (at least conceptually) dealt with on an equal footing and all of the states associated with different boundary conditions are figured into the enlarged Hilbert space from the beginning.

One of the main results of the present work is the boundary theory Lagrangian obtained by applying the techniques explored in \cite{Hatefi:2012bp,Park:2013iqa}, a variant of the standard Kaluza-Klein reduction. As we will see, Neumann-type boundary conditions will be crucial for the reduction ansatz that leads to the 3D theory. Having the explicit form of the 3D action
is quite advantageous in studying various aspects of 4D theory. 
The renewed understanding of the boundary conditions and dynamics offers a clearer picture of the black hole information paradox. As a matter of fact, such a step has been taken in \cite{Hatefi:2012bp}, \cite{Park:2017wiw} and \cite{Nurmagambetov:2018het}; the upshot of those investigations was that the boundary dynamics as well as the horizon deformation are the hair of the black hole. With the reduction to 3D and the explicit form of the action thereby obtained, this picture is made more concrete in this work.

\vspace{.2in}

The rest of the paper is arranged as follows. To be specific we take an Einstein-Hilbert action throughout. In section 2.1 and 2.2, we review the various boundary conditions, in particular the standard Gibbons-Hawking-York (GHY) term. The review of the GHY-term serves as a stage to engage various issues. Although the Dirichlet boundary condition has been widely used in gravitational and non-gravitational field theories, it has recently become evident that there is much more substance to the boundary conditions and dynamics. There are various ways of motivating Neumann or non-Dirichlet boundary conditions. Firstly and foremostly, the existence of the boundary dynamics itself is an indication of non-Dirichlet boundary conditions. They can also be motivated by the fact that in general the gauge transformations change the original boundary condition. 
If the initial boundary condition was a Dirichlet, all of the boundary conditions coming along with a coordinate transformation should be the Dirichlet in the new coordinate system, and we call them the ``Dirichlet-class" boundary conditions. However, from the viewpoint of the original coordinates, the boundary condition will not be that of the Dirichlet. 
To rephrase, the passive-viewpoint metric $g_{\m\n}'(x')$ will obey the Dirichlet for some appropriate hypersurface specified in terms of the $x'$ coordinate system, say, $x^{'3}=const$. Meanwhile the active-viewpoint metric $g_{\m\n}'(x)$ will satisfy a non-Dirichlet boundary condition. In section 2.3 we briefly comment on quantization and boundary conditions. In section 2.4 we present a generalization of the GHY-term for the Dirichlet boundary condition for a more general form of the action. Most of the analyses in section 2 can be carried out at a heuristic level. The more technical and in-depth issues are the main task of section 3 in which we carry out the Kaluza-Klein-type reduction, dimensional reduction to a hypersurface of foliation \cite{Park:2013vpa}, and obtain the 3D action. Having the explicit form of the 3D action puts one in a superior position to study the dynamics of the 4D theory through its connection to the 3D theory.
The association of the physical sector with the boundary \cite{Park:2014tia} makes it evident that the large gauge transformations (LGTs) become one of the main components of the system, and this is so even for the perturbative analysis: they will be directly relevant to the boundary degrees of freedom and their dynamics. 
After noting the well-known fact that a conformal Killing transformation transforms the metric by a scaling factor, we will see that the whole asymptotic conformal Killing symmetry \cite{Haco:2017ekf} - which contains the BMS symmetry \cite{Bondi:1962px}\cite{Sachs:1962wk}\footnote{The recent discussion of its roles in the infrared physics and black hole information can be found in \cite{He:2014laa,Hawking:2016sgy,Strominger:2017zoo}.} - is important for the 3D dynamics. Toward the end of section 3 we examine the insights offered by the 3D theory in black hole information (BHI). Being the hair of the black hole, the entire 3D dynamics will be responsible for BHI. In section 4 we have a summary and additional comments on the rationale and physical meaning for the enlarged Hilbert space. We end with future directions.
In Appendix we prove that the actively-transformed metric $g_{\m\n}'(x)$ satisfies the field equation. Although it may seem obvious, the procedure actually reveals some unexpected and interesting aspects of the gauge-fixing.


\section{Review of boundary conditions}

Although the Dirichlet boundary condition has been predominantly used in quantum field theories including gravitational theories (see, e.g., \cite{Parattu:2015gga}\cite{Lehner:2016vdi} for recent progress in the boundary conditions), the relevance of a Neumann-type (more generally, non-Dirichlet-type) boundary condition has become highlighted in some recent developments \cite{Park:2016fxc,Freidel:2016bxd,Krishnan:2016mcj,Krishnan:2016tqj}.\footnote{It is a bit ironic that the order of the developments of the two boundary conditions in string is the opposite: the Neumann boundary condition was exclusively used before the recognition of the importance of the Dirichlet boundary condition that led to D-branes.} One of the foci of the present work is the foliation of the spacetime and its implications on the boundary condition. The boundary conditions are crucially dependent upon the foliation, and there are various indications that boundary conditions other than the standard Dirichlet, in particular Neumann-type boundary conditions, must be considered with the Hilbert space enlarged accordingly - as stressed, e.g., in\cite{Freidel:2016bxd}. In this section we critically review the boundary conditions from the standpoint of the foliation: starting from the well-known Dirichlet condition, we raise various questions, and progress to the deeper aspects of the boundary condition and dynamics. Although the ideas involved should be valid quite generally, we often illustrate them with a Schwarzschild black hole to be specific and heuristic.

\subsection{Dirichlet boundary condition}

One encounters the issue of the boundary condition at several different stages. It first arises when applying the action principle to get the field equation(s) of the system.
To get to the bottom of the matter, we start by reviewing the variational procedure that led to the introduction of the GHY- term.
Let us consider an Einstein-Hilbert action with the GHY boundary term,
\be
S_{EH+GHY}\equiv S_{EH}+S_{GHY}  \la{awgh}
\ee
where
\be
S_{EH}\equiv \int d^4x \sqrt{-g}\; R\quad,\quad  S_{GHY}\equiv 2\int_{\pa {\cal V}} d^3x \sqrt{|h|}\, \ve K.
\la{ghyt}
\ee
For example, one can take the genuine time coordinate $t$ as the split direction, which is the usual foliation with $\ve=-1$. A different foliation and characterization of the boundary $\pa {\cal V}$ will be considered below.\footnote{ Later we will consider an $r$-slicing \cite{Papadimitriou:2004ap,Papadimitriou:2005ii,deBoer:1999tgo,Martelli:2002sp} with  $\ve=1$; the Dirichlet boundary condition in that case is along the $r$ direction.} The variation of $S_{EH}$ leads to (see, e.g., \cite{Padmanabhan:2014lwa} for more details)
\be
\!\!\!\!\d S_{EH}=\int_{\cal V} \sqrt{-g}\;G_{\m\n}\d g^{\m\n} -2\int_{\pa {\cal V}} d^3x\; \d(\sqrt{|h|}\, \ve K)\,   + \int_{\pa {\cal V}} d^3x\; \sqrt{|h|}\,\ve \Big(K h^{mn}-K^{mn}\Big)\d h_{mn}
\la{ehvv}
\ee
where 
\be
G_{\m\n}\equiv R_{\m\n}-\fr12 R g_{\m\n}.
\ee
The purpose of adding the GHY-term is to remove the second term above, thereby making it possible to get the Einstein equation upon the imposition of the Dirichlet boundary condition, $\d h_{mn}|_{\pa {\cal V}}=0$: the variation of the total action is 
\be
\d S_{EH+GHY}=\int_{\cal V} \sqrt{-g}\;G_{\m\n}\d g^{\m\n} +\int_{\pa {\cal V}} d^3x\; \sqrt{|h|}\,\ve \Big(K h^{mn}-K^{mn}\Big)\d h_{mn}.
\ee
The second term vanishes with the Dirichlet boundary condition, leading to the Einstein equation,
\be
G_{\m\n}=0.
\ee

The innocuous-looking analysis above actually harbors several potentially subtle issues that have their roots in the fact that the boundary conditions are imposed in conjunction with the foliation. Let us recall the general aspects of a symmetry: let $\Phi$ collectively represent the fields of the system under consideration, and consider a symmetry transformation. (For the present system, the field $\Phi$ is the metric and the symmetry transformation the diffeomorphism.)  
As well known, there exist two complementary viewpoints of the transformation: the passive viewpoint, $\Phi'(x')$, and the active viewpoint, $\Phi'(x)$.\footnote{{By definition, the active transformation is such that the function - in the present case, the field $\Phi$ - itself transforms to another function, $\Phi'$, without changing the argument $x$, whereas in the passive transformation, the coordinate $x$ does transform (with the corresponding transformation of the function as well).}} For a non-gravitational theory, which to choose is mostly a matter of convenience. However, in a gravitational theory, the subject becomes much richer and informative.
One way of seeing the need for various boundary conditions and thus the enlarged Hilbert space is through the gauge transformations that change the boundary condition: let us first consider the passive form of the general coordinate transformation:
\bea
g_{\m\n}'(x')=\fr{\pa x^\r}{\pa x'^\m}\fr{\pa x^\s}{\pa x'^\n} g_{\r\s}(x).
\eea 
In the new coordinates $x'^\m$, one will have a new time coordinate (as, e.g., in the Eddington-Finkelstein or null coordinates of a Schwarzschild geometry) and the foliation associated with the new time coordinate serving as the space of the leaves (i.e., the `base' in the fiber bundle parlance). Because of this the foliation content of the GHY-term is different from the original, although the form of the GHY-term remains the same. Two different solutions connected by a gauge transformation are considered equivalent; by the same token two boundary conditions would be considered the same if the hypersurface on which the boundary condition is imposed were kept the same (other than being expressed in the new coordinates). However, a natural boundary condition is normally associated with a different hypersurface, and because of this the boundary condition is changed. 
Let us illustrate this with a Schwarzschild black hole geometry,
\be
g_{S\m\n}(r,\th) dx^\m dx^\n \equiv  -\Big(1-\fr{2M}{r}\Big)dt^2+\Big(1-\fr{2M}{r}\Big)^{-1}dr^2+r^2 d\Omega^2. 
\la{ssm}
\ee
In the null coordinates,
\be
u\equiv t-r^*   \quad,\quad   v\equiv t+r^*
\ee
with
\be
r^* \equiv r+2M \ln\Big| \fr{r}{2M}-1\Big|,
\ee
the metric is written
\be
ds^2=-\Big(1-\fr{2M}r\Big)du dv+r^2 d\Omega^2.
\ee
In the null coordinates, the boundary hypersurface on which the boundary condition is imposed is changed due to the different foliation: a natural boundary condition is in terms of either $u$ or $v$ instead of $t$. In other words, one would choose $(u,r)$ or $(v,r)$ coordinates and impose the boundary conditions accordingly. (As we will elaborate later, a different foliation is associated with the observer-dependent effects at the quantum level.)

The fact that additional non-trivial issues are involved in the boundary conditions becomes evident by posing the following question: how do things look in the actively-transformed form of the  metric, $g_{\m\n}'(x)$? We now show that the passive transformation $g_{\m\n}'(x')$ and the active transformation $g_{\m\n}'(x)$ yield complementary pieces of information. The former satisfies the Dirichlet boundary condition in the new coordinates. However, the latter metric $g_{\m\n}'(x)$ - which can be interpreted as a ``new" solution\footnote{If one considered a large gauge transformation (LGT), $g_{\m\n}'(x)$ would represent a genuinely inequivalent solution. For an ordinary, i.e., a non-LGT, the matter is subtle; we will come back to this in section 3.2.3 where we elaborate on the effectively large gauge transformations.} in the original coordinates - satisfies a {\em non}-Dirichlet boundary condition in the original coordinate system $x^\m$. To see this, let us consider an infinitesimal 4D diffeomorphism\footnote{In Appendix we explicitly prove that $g_{\m\n}'(x)$ satisfies the field equation. The procedure reveals some unexpected and interesting aspects of the gauge-fixing.}: 
\be
g_{\m\n}'(x)= g_{\m\n}(x)+\nabla_\m \e_\n+\nabla_\n \e_\m  \la{ptm}
\ee
with a small parameter $\e_\m=\e_\m(t,r,\th,\f)$ that has, in particular, a non-trivial $t$-dependence.
Since the passively-transformed metric $g_{\m\n}'(x')$ satisfies the Dirichlet boundary condition in the new coordinate system $x'^\m$, such a transformation should definitely be allowed. In general, however, the active form $g_{\m\n}'(x)$ does not satisfy the Dirichlet boundary condition imposed in the original coordinates $x^\m$. Again this can be illustrated by taking $g_{\m\n}(x)$ to be a Schwarzschild black hole geometry $g_{S\m\n}(r,\th)$ in \rf{ssm}. Obviously, the infinitesimally transformed metric $g_{\m\n}'(x)\equiv g_{S\m\n}(r,\th)+{h}_{\m\n}(t,r,\th,\f)$ where ${h}_{\m\n} \equiv \nabla_\m \e_\n+\nabla_\n \e_\m$ does not satisfy the original Dirichlet boundary condition due to its time-dependence.

A different foliation can also be chosen within the given coordinate system. Let us consider the Schwarzschild spacetime. The standard Schwarzschild coordinates can be associated with the foliation where the $t$-direction is taken as the coordinate of the space of the leaves. Another possible slicing - which proves useful - is the $r$-foliation where the $r$-coordinate serves as the ``time" coordinate.  Although the $t$-splitting is useful for some purposes such as the non-perturbative bulk physics, the $r$-splitting should be a critical component in one's study of scattering around a black hole. With the standard GHY boundary term, but with the $r$-foliation, the boundary condition at $r=\infty$ is now that of Dirichlet except with $r$ playing the role of the ``time," i.e., ``$r$-Dirichlet." This boundary condition can be viewed as a Neumann-type \cite{Park:2017wiw}.

\subsection{Neumann and other boundary conditions}

In the previous subsection we saw that the description of the system from the active viewpoint leads to a Neumann type boundary condition.
As emphasized, e.g., in \cite{Freidel:2016bxd}, the heart of the matter is the observer-dependent effects in a gravitational theory: natural foliations and boundary conditions come with the coordinates system adapted to the reference frame of the observer. 
As will become clearer, the pertinence of different reference frames implies that the Hilbert space must be enlarged by incorporating the states of different boundary conditions. In this section we review the Neumann boundary condition that results from not adding the GHY-term; this Neumann boundary condition is different from that considered in section 2.2. (It will play a crucial role in section 3 where we work out the reduced action.) More generally, adding a different boundary term will lead to a different boundary condition just as the GHY- term leads to the standard Dirichlet boundary condition.

The Neumann-type boundary condition previously observed has arisen from considering a different foliation introduced by an active form of a gauge transformation within the same form of the GHY- term. A different-type, offshell-level, Neumann boundary condition is obtained by not adding (or more generally, modifying) the GHY- term\cite{Krishnan:2016mcj}\cite{Krishnan:2016tqj}. This Neumann-type boundary condition is within the context of the original foliation and is imposed in a manner analogous to imposing the Dirichlet with the standard GHY- term. To see this, let us revisit the variational procedure reviewed in the previous subsection.   
This time we employ the Hamilton-Jacobi procedure\cite{Sato:2002kv}\cite{Hatefi:2012bp}; the momentum is given by\footnote{
	The momentum in the Hamilton-Jacobi formalism is onshell; one intriguing feature of the present analysis is that the onshell momentum is associated with a 3D hypersurface since the bulk part of $\d S_{EH+GHY}$ vanishes onshell. It is also well known that the bulk ADM Hamiltonian vanishes and the ADM mass comes from the boundary terms. Therefore, there appear to be hints for the relevance and importance of the boundary degrees of freedom in several places in the classic analyses. In this respect it is interesting to note that in \cite{Regge:1974zd} the boundary conditions were viewed as canonical variables. 
}
\be
\pi^{mn}=\fr{\d S_{EH+GHY}}{\d h_{mn}}=\ve\sqrt{|h|} \Big(K h^{mn}-K^{mn}\Big). \la{hjm}
\ee
Note that to define the momentum field, the action $S_{EH+GHY}$, but not $S_{EH}$, is used; it is of course according to the standard practice. 
This expression of momentum implies that the boundary terms in \rf{ehvv} can be removed by requiring the following boundary condition
\be
 \d \pi^{mn}=0.
\ee 
In other words, variation of $S_{EH}$ leads, without the GHY-term (in 4D), to the field equation with the above Neumann-type boundary condition. Just as the Dirichlet boundary condition with the standard GHY- term represents a class of boundary conditions related by different foliations, the Neumann boundary condition above represents a class of boundary conditions which may be called the ``Neumann class."

As discussed in section 2.2, a change in the boundary condition can be caused by an ordinary (i.e., ``non-large") gauge transformation that introduces a different foliation within the same geometry.  The well-known example is the transformation between the Schwarzschild coordinates to the Kruskal coordinates. 
The change in the boundary condition can also be induced by different foliations within the same coordinate system (such as the $t$- vs $r$- foliation of a Schwarzschild geometry). As reviewed in the present section, a quite obvious way of changing the boundary condition is to add a different boundary term, and so far we have discussed the two most commonly used boundary conditions, the Dirichlet- and Neumann- classes.
As discussed in \cite{Chen:1998aw}, there should be many different types of boundary conditions (more on this in the conclusion) : adding various total derivative terms will induce the corresponding boundary condition changes at the offshell level.

\subsection{Quantization and boundary condition}

The quantization procedure imposes additional checkpoints on the boundary conditions. Consider the Einstein-Hilbert action. 
By shifting the metric according to 
\bea
g_{\m\n}\equiv {g}_{0\m\n}+ h_{\m\n}  \quad 
\eea
where $g_{0\m\n}$ denotes the background and $h_{\m\n}$ the fluctuation and adding a gauge-fixing term,
one gets the following kinetic terms:
\bea
\cL_{kin} = \sqrt{-g_0}\Big( -\fr12{\N}_\g h^{\a\b}{\N}^\g h_{\a\b}+\fr14 {\N}_\g h^{\a}_\a {\N}^\g h^{\b}_\b  \Big).
\eea
For the perturbative quantization one needs to compute the propagator. However, it was noticed long ago \cite{Kuchar:1970mu,Gibbons:1978ac,Mazur:1989by} that the path integral is not well defined due to the trace mode of the fluctuation metric $h_{\m\n}$. The problematic trace piece can be gauged away and the gauge-fixing procedure\footnote{The need for gauge-fixing of the trace piece is already revealed at the classical level as we note in the Appendix.} leads to the following constraint \cite{Park:2015xoa}\cite{Witten:2018lgb}: 
\be
K=K_0
\la{eq1q}
\ee
where $K$ and $K_0$ are the trace of the second fundamental form associated with $g_{\m\n}$ and $g_{0\m\n}$, respectively. 
As highlighted by Witten \cite{Witten:2018lgb},\footnote{The condition \rf{eq1q} with the boundary condition on the gauge-fixing term has been dubbed as the conformal boundary condition in \cite{Witten:2018lgb}.} the additional gauge-fixing above can be motivated from the fact that the Dirichlet boundary condition is not elliptic: the analysis of computing the propagator can be set up mathematically rigorously and only certain gauge-fixing is compatible with the existence of a propagator with the Dirichlet boundary condition.

In \cite{Witten:2018lgb} it was observed that the Dirichlet boundary condition does not lead to a well-defined perturbation expansion. This may be related to the finding in \cite{Park:2015xoa} that unless the trace part is fixed one gets a non-covariant result. In other words the reason that the usual Dirichlet boundary condition does not lead to a well-defined boundary value problems may have something do with the residual gauge symmetry\footnote{{In this respect it it interesting to note that a similar observation was made in \cite{Moss:2013vh}.}}: without the bulk gauge-fixing \rf{eq1q}, the kinetic operator does not define well-defined boundary value problem. It is presumably the infinite dimensional zero-modes \cite{Avramidi:1997sh,Witten:2018lgb} that cause (pre-loop) divergence, and manifests as non-covariance\cite{Park:2015ota}.

\subsection{Generalization of the Dirichlet GHY-term}

In spite of the newly-noticed relevance of the non-Dirichlet boundary conditions, one should not overlook the roles of the Dirichlet boundary condition. For instance the starting point of defining the canonical momentum is the action with the Dirichlet boundary condition. 
It is possible, as we show now, to extend the ``kinematics of Dirichlet vs. Neumann boundary conditions" to a more general system. Such an extension should be useful when dealing with the quantum-level action. Consider the following form of the action:
\be
S=\fr12 \int \sqrt{-g}\;f(R_{\m\n\r\s}) \la{sasakif}
\ee
where $f$ is an arbitrary function of the Riemann tensor. In place of \rf{sasakif}, one may consider the following first-order form of the action \cite{Deruelle:2009zk}:
\be
S=\fr12 \int \sqrt{-g}\;\Big[ f( \varrho_{\m\n\r\s}) +\vf^{\m\n\r\s}(R_{\m\n\r\s}-\varrho_{\m\n\r\s}) \Big] \la{1sto}
\ee
where $\varrho_{\m\n\r\s},\vf^{\m\n\r\s}$ are auxiliary fields. 
Let us consider the genuine time-splitting for simplicity; the spatial splitting case can be similarly analyzed. One can show that the 3+1 split form of the action \rf{1sto} is {\cite{Deruelle:2009zk}}
\be
S=\int  \Big[ \cL_{bulk} + \pa_\m (\sqrt{-g} \; n^\m K_{pq} \Psi^{pq})\Big]
\ee
where 
\bea
\cL_{bulk}&=& \sqrt{\g}N\Big[\fr12 f(\varrho_{\m\n\r\s})+\fr12 \phi^{ijkl}({R}_{ijkl}-{ \varrho_{ijkl}})-2\f^{mnp}(n^\k R_{mnp\k}-\r_{mnp}) \nn\\
&&\hspace{-.5in} -\Psi^{mn}\Big(KK_{mn}+K_{mp}K_n{}^p +n^{-1}D_m D_n n-\Omega_{mn}\Big)   -n^{-1}K_{mn} (\dot{\Psi}^{mn} -\mathscr{L}_{N^q\pa_q} \Psi^{mn} )
\Big].   \nn\\  \la{stoa}
\eea
In $\cL_{bulk}$ we have introduced
\bea
&&
\hspace{.6in}\quad  \r_{mnp}\equiv  n^\m\varrho_{mnp\m} \quad \Omega_{pq} \equiv n^\m n^\n\varrho_{p\m q\n} 
 \nn\\
&&\hspace{-.6in}\f^{mnpq}\equiv \g^{mm'}\g^{nn'}\g^{pp'}\g^{qq'}  \vf_{m'n'p'q'} \quad \f^{mnp}\equiv \g^{mm'}\g^{nn'}\g^{pp'} n^\m \vf_{m'n'p' \m} \nn\\
&& \hspace{1in}\Psi^{pq} \equiv \g^{pp'}\g^{qq'} n^\m n^\n \vf_{p' \m q' \n}.
\eea
By considering the 3D metric variation of \rf{stoa} and collecting the results one gets, for the momentum conjugate to $\g_{mn}$, 
\bea
p^{mn}&=& \sqrt{\g}\Big[-\fr12 \g^{mn} \Psi^{rs}K_{rs} -\fr12 \Psi^{mn}K -\Psi^{ml}K_l{}^n-\fr1{2n} (\dot{\Psi}^{mn}-\mathscr{L}_\b \Psi^{mn})     
    \nn\\
    &&+  \f^{mjnl}K_{jl}+\fr2{n} D_l (N\f^{lmn}) \Big]. \;\;
\eea
The trace part can be easily computed:
\bea
 p\equiv \g_{mn}p^{mn} &=& \sqrt{\g}\Big[-\fr52  \Psi^{rs}K_{rs} -\fr12 \Psi K -\fr1{2n} \g_{mn}(\dot{\Psi}^{mn}-\mathscr{L}_\b \Psi^{mn})
       \nn\\
       &&+ \g_{mn} \f^{mjnl}K_{jl}+\fr2{n} \g_{mn} D_l (N\f^{lmn})
\Big].   \;\;
\eea
Unlike the Einstein-Hilbert case, the present GHY-term, 
\be
S_{GHY}\equiv -\int \pa_\m (\sqrt{-g} \; n^\m K_{pq} \Psi^{pq}),
\ee 
and the boundary term that converts the Dirichlet action to the Neumann action (namely the Legendre transformation term) are different.(Recall that even in the Einstein-Hilbert case, they are different for $D\neq 4$ \cite{Krishnan:2016mcj}\cite{Krishnan:2016tqj}.)
Because of this, adding this term to the original action does not lead to the Dirichlet boundary condition. 
For the boundary Legendre transformation one can consider the variation of $S+S_{GHY}$ with respect to the 3D metric. This is just as in the Einstein-Hilbert case. The variation leads to the following boundary term:
\be
p^{mn}\d\g_{mn}.
\ee
To go to the action with the Neumann boundary condition, one should perform the Legendre transformation by adding the negative of $p$, $-p$, to $S+S_{GHY}$.

\section{Boundary dynamics}

Given the association of the physical states with the hypersurface at the boundary region \cite{Park:2015ybl} (which plays a central role in establishing the renormalizability of the physical states \cite{Park:2018vci} (and the refs therein)), a concrete understanding of the boundary dynamics and its coupling to the bulk is necessary for the complete picture. In the present section we apply a variant of the Kaluza-Klein dimensional reduction technique \cite{Hatefi:2012bp,Park:2013iqa,Park:2013vpa}, dimensional reduction to a hypersurface of foliation, and work out, up to some peculiarities, the Lagrangian of the 3D theory. We first carry out the reduction at the level of of an infinitesimal fluctuation and then discuss the full nonlinear extension.

Having the explicit form of the reduced Lagrangian, one can analyze various aspects of the original 4D system in a manner otherwise impossible. We first look, from the reduced theory's perspective, into the asymptotic symmetry aspect of the 4D theory by examining the conformal generalization \cite{Haco:2017ekf} of the BMS-type symmetry. Secondly, as we will observe, a Neumann-type boundary condition has an interesting implication for the Noether theorem. Lastly, we ponder the quantum aspects of the theory. 
The tie between the boundary condition and the foliation can be taken as an indication that the Hilbert space must include all those different foliation-induced boundary conditions.\footnote{The enlarged Hilbert space must also include the boundary-term-induced boundary conditions such as the Neumann-class; more on this in the conclusion.} The enlarged Hilbert space is important for black hole information as we will discuss toward the end.

\subsection{Projection onto holographic screen}

Although it was shown that the physical sector of the theory is associated with a 3D hypersurface in the boundary region,\footnote{As stressed in \cite{Park:2014tia}, the reduced theory is not a genuine 3D theory but still a 4D theory whose dynamics can be described through the hypersurface. For one thing, the graviton still has two degrees of freedom just as in a 4D theory.} the explicit form of the reduced action has not been obtained for a curved background. It can be obtained by  consistently reducing the 4D action; in what follows we will carry out the reduction in two steps: reduction of the 4D field equations to 3D and construction of the reduced action that reproduces the 3D field equations. 

Considering the (3+1) splitting:
\bea
x^\m\equiv (y^m,x^3), \quad  \m=0,..,3,\; m=0,1,2  \la{cs}
\eea
it is well known that the Einstein-Hilbert action
\bea
S_{EH}=\int d^4x \sqrt{-g}\;R  \la{eha}
\eea
can be cast into the ADM form \cite{Arnowitt:1962hi} (see \cite{Poisson} for a review)
\be
S_{EH} = \int_{\cal V}  d^4 x\;n\sqrt{|h|} \Big[\cR+K^2-K_{mn}K^{mn}\Big] -2\int_{\pa {\cal V}} d^3x \sqrt{|h|}\, \ve K. 
\la{admbulk}
\ee
The action contains a boundary term, the second term, in the ADM description. (The GHY-term, if added, cancels this boundary term.)
The second fundamental form $K_{mn}$ is given by 
\be
K_{mn}=\fr1{2n}\left(\mathscr{L}_{{3}} h_{mn}-{\nabla}_m N_{n}
-{\nabla}_n N_{ m} \right),\qquad K\equiv h^{mn}K_{mn}
\la{K4defqq}
\ee
{where $n$ and $N_m$ denote the lapse function and shift vector, respectively.} $\mathscr{L}_{{3}}$ denotes the Lie derivative along the vector field $\pa_{x^3}$ and $\N_m$ is the 3D covariant derivative.
We consider the 3+1 splitting where the $r$-direction is separated out. The $n,N_m,h_{mn}$ field equations are, respectively,
\be
\cR-K^2+K_{mn}K^{mn}=0  \la{ncon}
\ee
\be
{\N}_a (K^{ab}-h^{ab} K)=0  \la{Ncon}
\ee
\bea
&& \cR_{ab}-\fr12 \cR h_{ab} -\fr12 h_{ab}\Big[K^2-K_{pq}K^{pq}\Big] +   2KK_{ab}   -  2  K_{pa}h^{pq}K_{qb} \nn\\
&&+\fr1{n\sqrt{|h|}}h_{p a}h_{q b}\;\pa_r \Big[\sqrt{|h|}h^{pq} K  \Big]
-\fr1{n\sqrt{|h|}}h_{p a}h_{q b}\;\pa_r \Big[ \sqrt{|h|}K^{pq}  \Big] \nn\\
&&  -\fr2n h_{ab}\nabla _e (KN^e)  +\fr{2}{n}K \nabla_{(a} N_{b)}  
+ \fr2n \nabla^d (K_{ab}N_d) -\fr2n K_{n(a} \nabla^n N_{b)} =0  \nn\\
\la{metriceomNnz}  
\eea
where the symmetrization in \rf{metriceomNnz}, $(a \;b)$, is with a factor $\fr12$. 
The reduction procedure has two components. The first component is the requirement that the ansatze satisfy the 4D $n,N_m, h_{mn}$ field equations, \rf{ncon}-\rf{metriceomNnz}. As we will see, the requirement that the ansatze satisfy \rf{metriceomNnz} leads to the 3D version of the $h_{mn}$ field equation. The second component is to construct the ``3D" action that yields the 3D $h_{mn}$ field equation.
In section 3.1.1, we consider the reduction in static backgrounds including a Schwarzschild or Reissner-Nordstr\"{o}m black hole.
More general backgrounds including a Kerr black hole and even time-dependent black holes take additional care and are discussed in section 3.1.2.

\subsubsection{Static backgrounds}

Let us start with a static background metric of a diagonal form such as a Schwarzschild or Reissner-Nordstr\"{o}m geometry.
For such backgrounds a convenient gauge-fixing is to gauge away the fluctuation part of the lapse function and the shift vector $N_m$:
\bea
N_m=0\quad,\quad n=n_0(r)  \la{gfs}
\eea 
where $n_0(r)$ denotes the background solution for $n$.\footnote{One may take these gauge-fixings as part of the bulk Kaluza-Klein ansatze.} For a Schwarzschild background, for instance, it is 
\be
n_0=\left(1-\fr{2M}{r}\right)^{-\fr12}. 
\ee
With $N_m=0$, the $h^{ab}$ field equation \rf{metriceomNnz} becomes
\bea
&& \hspace{-.3in}\cR_{ab}-\fr12 \cR h_{ab} -\fr12 h_{ab}\Big[K^2-K_{pq}K^{pq}\Big] +   2KK_{ab}   -  2  K_{pa}h^{pq}K_{qb} \nn\\
&& \hspace{-.4in}+\fr1{n\sqrt{|h|}}h_{p a}h_{q b}\;\pa_r \Big[\sqrt{|h|}h^{pq} K  \Big]
-\fr1{n\sqrt{|h|}}h_{p a}h_{q b}\;\pa_r \Big[ \sqrt{|h|}K^{pq}  \Big] =0.
\la{metriceomq}  
\eea
As we will see shortly, the lapse function constraint \rf{ncon} is satisfied because it can be obtained by taking the trace of \rf{metriceomq}.  Substituting $N_{ a}=0$ into \rf{Ncon} one gets
\bea
{\N}^a \left[\fr{1}{n}\Big(\mathscr{L}_r h_{ab}
-h_{ab}h^{cd}\mathscr{L}_r h_{cd}\Big)\right]=0  \la{mtmconstr}
\eea
which is satisfied by the gauge-fixing above (more details can be found, e.g., in \cite{Park:2014tia}) 
\bea
\pa_a n=\pa_a n_0=0. \la{constronn}
\eea
It is therefore not necessary to further consider the shift vector constraint. As for the 3D hypersurface metric, let us first consider the linear-level reduction ansatz:
\be
h_{mn}(t,r,\th,\f) =h_{0mn} +\tilde{h}_{mn}(t,\th,\f) 
\la{ransq}
\ee
where $h_{0mn}$ denotes the solution of the field equation (for the Schwarzschild case, for instance, $h_{0mn}$ is given by $h_{0mn}=h_{0mn}(r,\th)=diag(n_0^2,r^2, r^2\sin^2\th)$)
and $\tilde{h}_{mn}(t,\th,\f)$ the fluctuation. We have shown in section 2 that such an ansatz must be allowed even though it does not obey the original Dirichlet boundary condition in the $(t,r,\th,\f)$ coordinates. 
Let us {\em choose} $\tilde{h}_{mn}(t,\th,\f)$ such that\footnote{Strictly speaking, this step is not necessary for finding the reduced form of the action. In other words, the fact that $\tilde{h}_{mn}=\nabla_m \e_n+\nabla_n \e_m$ satisfies the bulk and later the 3D field equations can be checked {\em after} the reduced action is obtained.}
\be
\tilde{h}_{mn}=\nabla_m \e_n+\nabla_n \e_m \la{hmnchoice}
\ee
for a parameter $\e_m=\e_m(t,\th,\f)$.
The ansatz \rf{ransq} is guaranteed to satisfy the $h_{mn}$ field equation \rf{metriceomq} for the following reason.\footnote{An explicit check is given in the Appendix. However, see the subtlety, therein explained, associated with gauge-fixing of the gauge parameter.} The right-hand side of \rf{hmnchoice} is nothing but the general coordinate transformation $\d h_{mn}$ with the gauge parameter $\e_m$.
In other words, $h_{0mn}(r) +\tilde{h}_{mn}(t,\th,\f)$ is guaranteed to be a solution of the $h_{mn}$ field equation since it takes a form identical to the gauge transformation of a solution $h_{0mn}(r)$. {(This also suggests how to obtain the nonlinear ansatz: just borrow the finite form of the gauge transformation of the background solution.)} 
A word of caution: we are {\em choosing} $\tilde{h}_{mn}$ {to be of the form \rf{hmnchoice}}; we are not generating $\tilde{h}_{mn}$ by utilizing the 3D gauge symmetry after starting with $h_{0mn} $, since the 3D symmetry is reserved for the gauge-fixing $N_m=0$. (For the same reason, one cannot gauge away $\tilde{h}_{mn}$.)

Finally, the trace part of \rf{metriceomq} is
\be
\fr12\Big(-\cR+K^2-K_{mn}^2 \Big)+\fr{h_{pq}}{n\sqrt{|h|}} \pa_r\Big( \sqrt{|h|}\, (K h^{pq}-K^{pq})\Big)=0.
\ee
{Consistency with \rf{ncon} requires that any solution of the bulk field equations \rf{ncon}-\rf{metriceomNnz} must therefore satisfy
\be
\fr{h_{pq}}{n\sqrt{|h|}} \pa_r\Big( \sqrt{|h|}\, [K h^{pq}-K^{pq}] \Big)=0. \la{hmntr}
\ee
}
{Since we are considering the ansatz given in \rf{ransq} that formally takes the form of the general coordinate transformation $\d h_{mn}$ with the gauge parameter $\e_m$, one may check \rf{hmntr} for the background $g_{0\m\n}$.} In other words, the ``gauge-transformed solution" \rf{ransq} is guaranteed to satisfy \rf{hmntr} (up to the subtlety discussed in the Appendix) and it is sufficient to check \rf{hmntr}  for the background $g_{0\m\n}$ to establish the consistency with \rf{ncon},
$-\cR+K^2-K_{mn}^2=0$.
This way, the lapse function constraint \rf{ncon} can be set aside, when constructing the reduced action, by the same token by which the shift vector constraint was previously set aside. {One can explicitly check that \rf{hmntr} is satisfied, e.g., by the Schwarzschild background. (Of course, eq. \rf{hmntr} will be generally satisfied by other backgrounds such as a Reissner-Nordstr\"{o}m solution.})

\vspace{.2in}
So far we have shown that the 4D field equations \rf{ncon}-\rf{metriceomNnz} are satisfied by our ansatze. 
The remaining task is to show that the field equation \rf{metriceomNnz} (or \rf{metriceomq}) - which is now viewed as a 3D field equation - can be derived from a the reduced action. In other words, the task is to construct the action of the 3D fluctuations whose field equation yields \rf{metriceomq}. 

Several remarks are in order before we get to the detailed steps of the construction. What we try to achieve is to work out the 3D Lagrangian that describes the physical {\em fluctuations} around a {\em given solution}. Let us use the 4D language for the moment. The fluctuations to be considered are the ones that would be generated by a gauge transformation if there were a residual symmetry: 
\be
g_{\m\n}'(x)= g_{0\m\n}(x)+h_{\m\n}(x)    \quad,\quad h_{\m\n} \equiv \nabla_\m \e_\n+\nabla_\n \e_\m  \la{ptmv}
\ee
where $g_{0\m\n}(x)$ and $h_{\m\n}(x)$ denote a given solution and the fluctuation, respectively.
Although the passively-transformed metric $g'_{\m\n}(x')$ satisfies a Dirichlet boundary condition, this will not be the case for the actively-transformed metric $g'_{\m\n}(x)$. This is because the background $g_{0\m\n}(x)$ satisfies the Dirichlet boundary condition whereas the fluctuation $h_{\m\n}(x)$ does not.  
Because of this it is not a priori clear whether one should start with an action with or without the GHY- term. It turns out, as we will soon see, that for consistent reduction one should use the action without the GHY- term,  \rf{admbulk}. Perhaps this is not entirely surprising since the reduced action is the action for the {\em fluctuation field} that satisfies the Neumann-type boundary condition.\footnote{It may be said that there exists a ``hierarchy" in the boundary conditions: a solution is found usually by imposing a Dirichlet boundary condition. (In general, there may well be solutions that satisfy a different boundary condition such as a Neumann.) For the fluctuations around the solution, there will be different sectors that satisfy different boundary conditions. In particular there will be a sector that satisfies a Neumann-type boundary condition. As discussed in section 2 such a sector must be included because it is obtained by a gauge transformation (which effectively acts like an LGT; see section 3.2.3). One subtle point here is that the Neumann boundary condition that plays an important role below in constructing the reduced action is that of the Neumann class, namely the boundary term-induced one discussed in section 2.2. Meanwhile it was the Dirichlet-class Neumann boundary condition, namely the foliation-induced Neumann boundary condition, that has motivated the ansatz \rf{hmnchoice} and provided the rationale for the enlarged Hilbert space. We will come back to this in the conclusion.}
In other words, the 3D action describes the boundary theory whose dynamics is what makes the boundary condition - from the bulk point of view - deviate from the Dirichlet. Since the reduction is carried out in the original coordinates, the action without the GHY-term may somehow become relevant. Indeed, this is what happens.

Let us now get to the construction of the 3D action. The derivation of the field equation \rf{metriceomq} by starting with the 4D action \rf{admbulk} involves partial integrations along $r$. We will now show that the form of the sought-after 3D action is what one inherits from the 4D action $S_{EH}$, the action without the GHY-term: the field equation \rf{metriceomq} is obtained from that 3D action {\em without} performing the partial integration along $r$.
As we have reviewed in section 2, the Neumann boundary condition must be imposed for the variational procedure without the GHY-term. In this sense it can be said that the Neumann boundary condition discussed in section 2.2 makes it possible to get eq. \rf{metriceomq}. 
In other words, if one starts with \rf{admbulk}, which is the action without the GHY-term, and takes the $h_{mn}$ variation, one gets \rf{metriceomq} without performing the $r$-partial integration. We now show this by repeating the derivation of the field equation in the 3D setup. Consider the following 3D action
\be
S_{reduced} = \int d^3x\;n_0 \sqrt{|h|} \Big[\cR+K^2-K_{mn}K^{mn}\Big]-2 \int d^3x \sqrt{|h|}\; \ve K
\la{ehv}
\ee
where we have set, i.e., gauge-fix, $n=n_0$ and $N_m=0$; see the statements around \rf{gfs}. (Due to the  $N_m=0$ gauge-fixing, for example, the shift vector-containing terms in the definition of $K_{mn}$ (and $K$) in \rf{K4defqq} are now absent.) The first term has been inherited from the bulk term of the 4D action and the second term inherited from the boundary term. Recall that the boundary term in \rf{admbulk} is {\em not} itself the GHY- term but negative of that: if the GHY- term were added, it would cancel out this boundary term. {Below we will show, up to some peculiarities, that this action with the Neumann boundary condition reproduces eq. \rf{metriceomq} (without the second line for the reason to be explained).} 

{A word of clarification is in order. What we are up to is establishment of \rf{ehv} as the sought-after reduced action. One may wonder in what sense it is a ``reduced" action. To recognize \rf{ehv} as a reduced action, let us summarize what we have shown above. After starting with the bulk field equations \rf{ncon}-\rf{metriceomNnz}, we have made the following {\em reduction ansatze}:
\bea
&& \hspace{.3in}g_{3m}=0,\quad g_{33}=g_{033}  \nn\\
&& g_{mn}(t,r,\th,\f) =h_{0mn} +\tilde{h}_{mn}(t,\th,\f) 
\eea
where in the first line we have deliberately used the original metric variables $g_{\m\n}$ instead of the ADM variables $N_m,n, h_{mn}$\footnote{
Recall that
\[
g_{\m\n}=\left(
\begin{array}{cc}
h_{mn} & N_{ m} \\
&\\
N_{ n} &  n^2+h_{mn}N^{m} N^{ n} 
\end{array}
\right).
\]
} to make it clearer that the above can be viewed as the Kaluza-Klein reduction ansatze. Although the first line was previously viewed as gauge-fixing, it can also be viewed as part of the Kaluza-Klein reduction ansatze. As shown above, it is no longer necessary to consider the $n, N_m$ field equations, and this is part of the reason that the action \rf{ehv} can be viewed as reduced. The hallmark of the standard Kaluza-Klein procedure is that one gets the field equation(s) of the lower dimensional fields (presently, $\tilde{h}_{mn}(t,\th,\f)$) and the consistency of the reduction is established once the lower dimensional action - which reproduces the lower-dimensional field equations - is obtained. The reduced action describes the dynamics of the lower-dimensional field(s) - presently, $\tilde{h}_{mn}(t,\th,\f)$ - that obviously do not depend on the $r$-coordinate. 
}

The explicit steps to lead to the 3D version of the $h_{mn}$ field equation are as follows. 
The variation of the first term in \rf{ehv} is
\bea
&&\int d^3x\, {n_0}\sqrt{|h|}\,\cR_{ab}\, \d h^{ab}-\fr12 {n_0}\sqrt{|h|} \;h_{ab}  \Big[\cR+K^2-K_{mn}K^{mn}\Big]\d h^{ab} \nn\\
&&+   2 { n_0} \sqrt{|h|}\,(KK_{ab}-   K_{pa}h^{pq}K_{qb})\d h^{ab}
+ 2{ n_0}\sqrt{|h|}\, \Big[K  h^{ab}  \d K_{ab}- K^{pq}\d K_{pq} \Big]   \nn\\  \la{bulkvari}
\eea
{where the variation is $\d h_{pq}=\d \tilde{h}_{pq}$ (since we are considering $\d \tilde{h}_{pq}$, really, through a ``chain rule").}
One can rewrite the fourth term as
\bea
&& 
\sqrt{|h|}\, \Big[K h^{ab} \d\pa_r h_{ab}- K^{pq}\d \pa_r h_{pq} \Big] 
=\sqrt{|h|}\, \Big[K h^{ab} - K^{pq}\Big] \d\pa_r h_{ab}=\pi^{ab} \d\pa_r h_{ab}  \nn\\
&&{ =\pi^{ab} \pa_r \d h_{ab}=\pi^{ab} \pa_r \d \tilde{h}_{ab}=0} \la{ibdterms}
\eea
where $\ve$ in \rf{hjm} has been set to $\ve=1$ - because $r$-foliation is being considered - in the second equality.
{The last equality follows from the fact that $\d \tilde{h}_{ab}$ (and of course $\tilde{h}_{ab}$) is $r$-independent.} The variation of the second term in \rf{ehv} is
\bea 
&& -2\d \int_{\pa {\cal V}} d^3x \sqrt{|h|}\,  K= -\d \int_{\pa {\cal V}} d^3x  h_{ab}\pi^{ab} = -\d \int_{ {\cal V}} d^4x \pa_r( h_{ab}\pi^{ab})   \nn\\
&&=- \int_{ {\cal V}} d^4x \pa_r\Big[(\d h_{ab})\pi^{ab} + h_{ab}\d\pi^{ab} \Big] 
\eea
where we have used the trace of \rf{hjm} in the first equality.
The second term in the second line goes
\bea
- \int_{ {\cal V}} d^4x \pa_r\Big[h_{ab}\d\pi^{ab} \Big] =- \int_{\pa {\cal V}} d^3x \Big[h_{ab}\d\pi^{ab} \Big]
\eea
and thus vanishes upon imposing the Neumann boundary condition $\d\pi^{ab}=0$.
The first term in the 2nd line takes
\bea
- \int_{ {\cal V}} d^4x \Big[ (\pa_r \d h_{ab})\pi^{ab}+ \d h_{ab}\pa_r \pi^{ab}\Big].
\eea
This with \rf{ibdterms} leads to:
\bea
- \int_{ {\cal V}} d^4x \, (\d h_{ab}) \pa_r \pi^{ab}.
\eea
{Since the $r$-Dirichlet boundary condition $\d\pi^{ab}=0$ is compatible with the condition $\pa_r \pi^{ab}=0$ (an analogous statement for a genuine time case can be found, e.g., in \cite{Polchinski:1996na})\footnote{One subtlety is that the condition $\pa_r \pi^{ab}=0$ is not satisfied, e.g., by a Schwarzschild solution. It appears that for a more rigorous analysis, it is necessary to include an extra boundary term (also, see footnote 24), a Neumann-analogue of $S_0$ that is present in eq. (4.7) of \cite{Poisson}. Let us quote (4.7)-(4.10) therein: $S_G=S_H+S_B-S_0$
where
\[
S_H=\fr1{16\pi}\int_{{\cal V}} d^4x \sqrt{-g}\;R \nn\\
\]
\[
S_B=\fr1{8\pi} \int_{\pa{\cal V}} d^3y \sqrt{|h|}\; \ve K
\]
\[
S_0=\fr1{8\pi} \int_{\pa{\cal V}} d^3y \sqrt{|h|}\; \ve K_0.
\]
$K_0$ is interpreted as the extrinsic curvature of the boundary embedded in flat spacetime. More generally, one may consider $K_0$ as the background extrinsic curvature. More generally, the form of the term analogous to $S_0$ should depend on the boundary condition. In this spirit, for the present case, adding (with an appropriate sign and factor) the following term to \rf{ehv} will do the job, i.e., remove the aforementioned non-vanishing contribution of the background to $\pa_r \pi^{ab}$, 
\[
S_0\sim \int d^3x h_{ab}\pi_0^{ab}\sim  \int d^3x \sqrt{|h_0|}\;h_{ab}K_0^{ab}.
\]
Also, there may well be solutions that do satisfy the condition $\pa_r \pi^{ab}=0$; for them such extra device will not be necessary.
}, combining all the results above and using the expression for the momentum field \rf{hjm} reproduces \rf{metriceomq} without the second line (which vanishes due to the Neumann boundary condition $\pa_r \pi^{ab}=0$), as promised.}

\subsubsection{General backgrounds}

Some of the steps above need to be modified before application to a more general background. Here we present an approach that should be applicable even to time-dependent backgrounds. To see why some of the steps above need modifications, let us take the Kerr case to be specific and contrast it with the Schwarzschild case. The difference between these two cases is the manner in which the gauge-fixings of the lapse function and shift vector satisfy the shift vector constraint \rf{Ncon}, which we quote below for convenience, 
\be
{\N}_a K^{ab}-h^{ab} {\N}_a K=0.  \la{Nconq}
\ee
While the gauge-fixing $N_m=0, n^2=n_0^2(r)=\left(1-{2M}/{r}\right)^{-1}$ for the Schwarzschild case makes each term in {\rf{Nconq}} separately vanish, the same is not true for the Kerr case. For a Kerr metric 
\be
g_{0\m\n}=\left(
\begin{array}{cccc}
 \frac{2 M r}{\r^2}-1 & 0 & 0 & -\frac{2 a M r \sin ^2\theta }{\r^2} \\
 0 & \frac{\r^2}{\D} & 0 & 0 \\
 0 & 0 & \r^2 & 0 \\
 -\frac{2 a M r \sin ^2\theta }{\r^2} & 0 & 0 & \frac{\Sigma \sin ^2\theta }{\r^2} \\
\end{array}
\right)
\ee
where 
\be
\r^2=r^2+a^2\cos^2\th\quad,\quad \D=r^2-2Mr+a^2
\quad,\quad \Sigma=(r^2+a^2)^2- a^2\D \sin^2\th
\ee
one can adopt the analogous gauge-fixings: 
\be
N_m=0, n^2=n_0^2(r,\th)=\fr{\r^2}{\D}.
\ee
It is not difficult to show that the shift vector constraint \rf{Nconq} is satisfied at the $\tilde{h}_{mn}$-linear order.
At the $\tilde{h}_{mn}$-linear order: \rf{Nconq} becomes the leading field equation which is of course satisfied by the Kerr background; the reduced action is  again given by \rf{ehv}. However, the $\tilde{h}_{mn}$-higher order status is not so obvious.
Moreover, for a more general background (such as a time-dependent background) a similar gauge-fixing that solves the constraint may not be available and it may be necessary to keep the shift vector instead of setting it to a fixed value. Keeping $N_m$, the reduced action is again given by 
\be
S_{reduced} = \int d^3 x\;n_0 \sqrt{|h|} \Big[\cR+K^2-K_{mn}K^{mn}\Big]-2 \int_{\pa {\cal V}} d^3x \sqrt{|h|}\; \ve K
\la{ehvN}
\ee 
which has the same form as \rf{ehv}. However, here, $N_m$ is nonzero and acts as a Lagrange multiplier. This reduction procedure should cover quite general backgrounds including the aforementioned time-dependent ones.

\subsection{Symmetry {implications} of the 3D theory}

In this section, we utilize \rf{ehv} (or \rf{ehvN}) to study the symmetry aspects of the original 4D theory. In particular, we examine the physical meanings of the BMS symmetry by starting with the asymptotic conformal Killing symmetry or the conformal BMS group \cite{Haco:2017ekf} that contains the BMS group as a subgroup. The conformal BMS symmetry and its present analogue have an intuitively clear meaning and thus further clarify the meaning of the BMS group.

\subsubsection{Reflection on BMS}

Let us recall the generalities on an unbroken symmetry. Let us denote, as before, the field of the system under consideration by $\Phi$ and imagine splitting the field into a fixed background $\Phi_0$ and the fluctuation $\tilde{\Phi}$:
\be
\Phi=\Phi_0+\tilde{\Phi}.    \la{gex}
\ee
The symmetry group of the theory gets {``spontaneously"} broken into a subgroup that leaves $\Phi_0$ invariant. 
The BMS symmetry is unbroken in the above sense, except that  it is not a precise symmetry but an asymptotic symmetry {of $\Phi_0$}. The BMS symmetry has been extended in \cite{Haco:2017ekf} to the so-called conformal BMS group, the {\em asymptotic} conformal Killing symmetry. 

{Let us first review the conformal Killing symmetry itself before exploring its meaning in the present context.} 
The diffeomorphism contains a particular form of the conformal transformation
as can be seen by rewriting the diffeomorphism transformation with a parameter $\e^\m$ in \rf{ptm} as
\bea
\d g_{\m\n}=\fr12 (\nabla_\k \e^\k)g_{\m\n}+(Lg)_{\m\n} \la{diffeo}
\eea
where 
\bea
(Lg)_{\m\n}\equiv \nabla_\m \e_\n+\nabla_\n \e_\m-\fr12 (\nabla_\k \e^\k)g_{\m\n}.
\eea
Note that the first term of \rf{diffeo} takes the form of a conformal transformation. Suppose for the moment that $\e_\n$ is a precise (i.e., not just asymptotic) conformal Killing vector, that is, it satisfies
\be
\nabla_\m \e_\n+\nabla_\n \e_\m=\fr12 (\nabla_\k \e^\k)g_{\m\n}.   \la{ckv}
\ee
For this particular $\e^\m$, the diffeomorphism acts as a conformal-type transformation since $(Lg)_{\m\n}=0$:
\bea
\d g_{\m\n}=\fr12 (\nabla_\k \e^\k)g_{\m\n}.    \la{diffeocon}
\eea
The asymptotic conformal Killing symmetry is a symmetry generated by $\e^\m$ that satisfies \rf{ckv} not precisely but asymptotically \cite{Haco:2017ekf}.

Let us examine the physical meaning of an asymptotic symmetry in the present context, which is analogous to the conformal BMS group. As briefly stated below \rf{gex}, for an exact symmetry, one would consider 
\bea
g_{\m\n}\equiv  {g}_{0\m\n} +h_{\m\n} 
\eea
and look for a subgroup that leaves the background ${g}_{0\m\n}$ invariant. Compared with this, there are several things that make the analysis of an asymptotic symmetry more unwieldy than otherwise. The first and most obvious is the fact that the symmetry under consideration is not an exact symmetry but an asymptotic one.\footnote{In the review by Strominger \cite{Strominger:2017zoo}, the asymptotic symmetry is identified with the LGT (see eq. (2.10.1) therein). The definition that we adopt is more general in the sense that the asymptotic symmetry includes the symmetry that leaves the boundary invariant but not the bulk. (It was also noted in \cite{Strominger:2017zoo} that there are cases where the definition given in (2.10.1) is too narrow.)} If the asymptotic conformal Killing symmetry were an exact symmetry then it (and therefore the BMS symmetry) would be the unbroken symmetry. Since it is an asymptotic symmetry, it may be called the ``asymptotic unbroken symmetry."
The second thing is the fact that the asymptotic region considered in \cite{Haco:2017ekf} was the null infinity, which is different from the boundary at an spatial infinity obtained, e.g., by taking $r\ra \infty$ as in the present work.\footnote{See, e.g., \cite{Sommers} for a discussion of such an asymptotic region.} As often demonstrated in the literature, however, different asymptotic regions usually have corresponding quantities \cite{Ashtekar:1978zz}.\footnote{Roughly speaking, we are considering the conformal extension of the asymptotic group of \cite{Ashtekar:1978zz}.} The third thing is the fact that the ADM formalism makes it less transparent to apply the results of \cite{Haco:2017ekf} to the present setup. Because of these reasons, not all of the statements below are entirely rigorous (and we do not attempt to make them more rigorous in the present work). Nevertheless, we present the following picture for its enlightening perspective. 

Let us examine the symmetry aspect for the reduced theory - which we quote here for convenience: 
\be
S_{reduced} = \int d^4 x\;n_0 \sqrt{|h|} \Big[\cR+K^2-K_{mn}K^{mn}\Big]-2 \int_{\pa {\cal V}} d^3x \sqrt{|h|}\; \ve K.
\la{ehvq}
\ee
Now it is to be understood that 
\be
h_{mn}(t,r,\th,\f) =h_{0mn} +\tilde{h}_{mn}(t,\th,\f) 
\ee
is substituted for $h_{mn}(t,r,\th,\f) $. For simplicity we again consider the infinitesimal fluctuation case.
The conformal Killing group will act as the symmetry of the boundary theory. As previously stated, this is not an exact symmetry of the bulk theory in the usual sense but {presumably the closest} analogy one can get for the asymptotic conformal Killing group. The physical meaning that we are after is the role played by the symmetry in the 3D boundary theory: the symmetry will generate a set of inequivalent vacua, which will be an important part of the 3D description of the 4D dynamics. The 3D Fock space will then be built on these {inequivalent} vacua.   
What is rather surprising is that all those different Fock states will have the common bulk configuration $g_{0\m\n}$; the 3D dynamics will be important for black hole information as we discuss in section 3.2.4.

\subsubsection{Noether charge-related}

As we already saw, there are several indications of the necessity of the Neumann-type boundary conditions: for instance, if one tries to describe the dynamics from the active coordinate transformation, one encounters such a boundary condition.\footnote{There also is an indication at the quantum level: the works of \cite{Park:2016vam} and \cite{Nurmagambetov:2018het} show that having a Dirichlet boundary condition is classically sufficient, whereas generically, a Neumann-type or non-Dirichlet boundary condition is required at the quantum level.} Here we examine the implication of the foliation-induced (i.e., the Dirichlet-class) Neumann boundary condition for the Noether theorem.

Essentially, the implication is that the Noether current that used to be conserved in a setup with the Dirichlet boundary condition is no longer conserved in a setup with the Neumann boundary condition. The implication seems useful for understanding certain aspects of the conserved quantities (or quantities viewed as conserved in the conventional description with the Dirichlet boundary conditions) of a black hole through quantum effects such as the Hawking radiation. Let us illustrate this with the mass or entropy of a black hole. With the Hawking radiation, the mass and entropy of the black hole will decrease. This seems incompatible with a non-dynamic boundary, i.e., a boundary with a Dirichlet boundary condition since a non-dynamic boundary would imply conservation of the charges.
The mass or entropy decrease must have something to do with a Neumann-type boundary condition \cite{Park:2017wiw} since such a boundary condition would imply non-conservation of the mass or entropy. 

To deliver the punch line with minimum complications, let us first consider a non-gravitational system in a flat background and briefly review the Noether theorem. Suppose the system whose field is $\Phi$ has a global symmetry:  
\be
\Phi \ra \Phi+\e \d \Phi.
\ee
Let us now make the parameter $\e$ local.
On general grounds, the variation of the action must take the following form:
\be
\d S=\int J^\m \pa_\m \e
\ee
where $J^\m$ is the Noether current.
If one takes $\Phi$ as a solution of the field equation, the action must be stationary and thus 
\be
\d S=0. 
\ee
Suppose the field $\Phi$ and its variation $\d \Phi$ satisfies the Dirichlet boundary condition as normally assumed. Then the two equations above imply
\be
\pa_\m J^\m=0 \la{ucc}
\ee
and this is the standard current conservation law which in turn leads to the charge conservation. 
For a Neumann boundary condition, the boundary term does not vanish and one gets 
\be
\d S=\int_{ {\cal V}} J^\m \pa_\m \e=\int_{ \pa{\cal V}} n_\m J^\m  \e-\int_{ {\cal V}}  \e \pa_\m J^\m =0
\ee
which, instead of \rf{ucc}, implies
\be
\int_{ {\cal V}}  (\pa_\m J^\m) \e = \int_{ \pa{\cal V}} (n_\m J^\m)  \e.
\ee
This simply means that the bulk current is not conserved but coupled with the corresponding boundary quantity.

Let us turn to the Einstein-Hilbert case. We consider the $t$-foliation and the foliation-induced Neumann boundary condition. From the fact that the diffeomorphism variation (under $x^\m \ra x'^\m=x^\m-\xi^\m$) is essentially a Lie dragging, it follows that
\be
\d_{\xi} (\sqrt{-g}\; L)= \sqrt{-g}\; \nabla_\m (\xi^\m L). \la{ldrag}
\ee
Meanwhile the diffeomorphism variation $\d_{\xi}$ is a special case of an arbitrary variation $\d$ and thus
\bea
\d_{\xi} S_{EH}&=&\int_{\cal V} \sqrt{-g}\;\Big[G_{\m\n}\;\d_{\xi} g^{\m\n} +\nabla^\r v_\r \Big]
\la{ehvdiffN}
\eea
where
\be
v_\r=\nabla^\l \d g_{\r\l}-g^{\a\b}\nabla_\r \d g_{\a\b}.
\ee
The equivalence of \rf{ldrag} and \rf{ehvdiffN} leads to the current $J^\m$
\be
J^\m\equiv -2G^{\m\n}\xi_\n+v^\m-L\xi^\m
\ee
with
\be
\nabla_\m J^\m=0.
\ee
Let us show that the current associated with the ``new solution" with the Neumann boundary condition, given in \rf{ptm} (here we use $\xi^\m$ instead of $\e^\m$), does not satisfy the current conservation. (This of course implies the charge (i.e., mass or entropy) is not conserved.) To see this, consider the active transformation of the current
\be
J'^\m= J^\m+\d J^\m= J^\m+ \mathscr{L}_\xi J^\m
\ee
and the volume integral of its divergence: 
\be
\int d^4x \sqrt{-g}\; \nabla_\m {J'}^\m=\int d^4x \sqrt{-g}\; \nabla_\m   \mathscr{L}_\xi J^\m     
\ee
where $\nabla_\m J^\m=0$ has been used to obtain the right-hand side. By applying the Stokes' theorem one gets
\be
\int d\Sigma_\a\;    \mathscr{L}_\xi J^\a |    
\ee
where the vertical line `$|$' indicates that the integrand be evaluated at the boundary $\Sigma$. The expression above vanishes for a Dirichlet class since $\xi=0$ at the boundary. However, the same is not true for the $\xi^\m$ that satisfies the Neumann boundary condition: the mass or entropy decrease via Hawking radiation is connected with the Neumann boundary condition. We will come back to this point in section 3.2.4 below.

\subsubsection{Effectively large gauge transformations}

In section 2.1 we discussed that in the active (as opposed to passive) view a change in the boundary condition can be caused by a gauge transformation of the metric, $g'_{\m\n}(x)$. If the gauge transformation is a large one, the transformed metric represents an inequivalent solution.
A non-LGT is usually viewed to generate a gauge-equivalent solution. As we discuss now, this matter is not so clear-cut once one considers an asymptotic region. The idea can be illustrated with a Schwarzschild black hole geometry; consider the null coordinates 
\be
u\equiv t-r^*   \quad,\quad   v\equiv t+r^*
\ee
with
\be
r^* \equiv r+2\a M \ln\Big| \fr{r}{2M}-1\Big|
\ee
where we have inserted a small parameter $\a$ for convenience. 
This transformation between $(t,r)$ and $(u,v)$ can be viewed as small only for a finite $r$. Once one considers an asymptotic region $r=\infty$, however, any finite small parameter $\a$ will not make the transformation `small.' In this sense it can be said that at $r=\infty$ the transformation is effectively an LGT. As discussed in section 2, such a transformation is relevant for the boundary dynamics: the transformed solution acts as a new solution just as an LGT-generated solution is an inequivalent one.

\subsubsection{Quantum effects and BHI}

There are several facets of the quantum-level dynamics and boundary conditions; we focus on the aspects relevant to the black hole information problem. 
There has been a proposal in loop quantum gravity that the Hilbert space must be enlarged to include all those states associated with the `extended Gibbons-Hawking' boundary term \cite{Freidel:2016bxd}. What we observe in the present work is in line with the proposal, and consideration of the enlarged Hilbert space must be a necessary condition for solving the information problem.

In the previous sections we have discussed several rationales for the enlarged Hilbert space from the standpoint of the foliation-based quantization of gravity. One of them was the boundary condition-changing gauge transformations. 
A change between the reference frames with the accompanying transformation between the adapted coordinate systems brings observer-dependent effects. This is well known, e.g., in the descriptions of a quantized scalar field in a Schwarzschild black hole background by employing Schwarzschild and Kruskal coordinates \cite{Birrell}. Each coordinate system has the associated vacuum: the Schwarzschild vacuum (Boulware vacuum) and the Kruskal vacuum (Hartle-Hawking vacuum). The Kruskal vacuum appears to a Schwarzschild observer as thermally radiating. The presence of such inequivalent vacua is an essential part of the setup that ultimately leads to the black hole information paradox. 
By the same token, the BMS transformations introduce many different inequivalent vacua and the BMS charges or their conformal extension will be observable to a Schwarzschild observer. In each of those vacua, it will be possible to perform the transformation between Schwarzschild and Kruskal coordinate systems; the transitions between all these different vacua must be of the information-minimal type \cite{Park:2017wiw}. The information-carrying gravitons must be the ones that are associated with the 3D fluctuations.

\section{Conclusion}

In this work we have reviewed boundary conditions and examined various pertinent issues. The present further scrutiny of the widely-used Dirichlet boundary conditions has been motivated by the fact that the physical sector of a gravity theory is associated with the hypersurface at the boundary.
We have stressed that the boundary condition is closely tied with the foliation of the spacetime. This implies that the effects of the gauge symmetry on the boundary condition must be carefully tracked.   
It also implies that for the given boundary terms added to the action, such as the GHY-term or its variations, {there corresponds a class of the boundary conditions.} For instance, there exist Dirichlet- or Neumann- class boundary conditions. 

When a gauge transformation is examined in the active viewpoint, the need for an enlarged Hilbert space with various different boundary conditions naturally arises: a gauge transformation, ordinary or large, does not in general preserve the boundary condition originally imposed. In addition to the Dirichlet-class Neumann boundary condition, there is a Neumann-class boundary condition that one obtains by not adding, in 4D, the GHY-term.    
The Neumann-class boundary condition plays a crucial role in reducing the bulk theory to a 3D theory\footnote{In our previous works \cite{Park:2013iqa,Park:2013vpa,Park:2013bma} where dimensional reduction to a hypersurface of foliation was discussed, the consistency of the reduction required certain boundary terms. \nolinebreak This might have the same origin as the omission of the GHY-term and need of an extra boundary terms discussed in section 3.1.}  whose explicit form has been obtained in section 3. 
With the 3D action available we have studied various aspects of the 4D theory. In particular, the conformal extension of the BMS type symmetry - and thus the BMS-type symmetry - comes to have a clearer meaning in the 3D description of the 4D  physics. We have confirmed and made more concrete our earlier proposal that the 3D dynamics is a critical part of the black hole hair.

For further directions, it may be useful first to know where the present work stands in relation to our recent sequels. 
It was realized in \cite{Park:2014tia,Park:2015xoa,Park:2015ybl} that a large amount of the diffeomorphism can be tamed in such a way to lead to the projection of the physical states of the theory on to the holographic screen in the asymptotic region. 
Although the offshell non-renormalizability was established in the past, which was confirmed by the results in our recent works, renormalizability of the physical sector can be established on the aforementioned reduction of the physical states. The renormalization procedure requires perturbative computation of the 1PI effective action by employing refined background field method \cite{Park:2013vpa}. The reduction of the physical sector is what makes it possible to realize 't Hooft's idea on the renormalizability based on the metric field redefinition. The one-loop renormalization has been carried out for pure gravity, a gravity-scalar system, and an Einstein-Maxwell theory \cite{Park:2018vci} (and references therein). It would have been more desirable to discuss the issues of the boundary conditions prior to these works. This is especially true because in obtaining the 1PI action and the quantum-level field equations (which were used in some of the subsequent applications of the results), more proper treatment of the boundary conditions is necessary. Although the lack of the explicit form of the reduced action was not a fundamental obstacle to establishing the renormalizability, having this form gives one advantages in studying various other aspects of the theory. As we have discussed in section 3, one such example is the appreciation of the conformal extension of the BMS symmetry, the BMS symmetry itself, and their roles in the 3D description of the bulk theory; the conformal BMS-type symmetry is one of the crucial components of the 3D theory. Another advantage of having the explicit form of the 3D theory is the insights that it offers on the black hole information paradox. Our picture makes it clear that the 3D dynamics is a crucial component of the system information.

\vspace{.3in}
The status quo above suggests several further directions.
In the main body we have classified the boundary conditions into three different categories: the Dirichlet class (section 2.1), the Neumann class, and more general boundary conditions (section 2.2). The recent developments seem to indicate that all of these sectors should be included in an enlarged Hilbert space. Since the bulk configurations are all common, all of the different boundary conditions must be associated with different aspects of the bulk fluctuation {\em reflected on the boundary}. It will be interesting to explore whether it is possible (and if so, how) to impose the most general boundary condition inclusive of all the possible boundary conditions. 
It will also be of some interest to explore in more detail the enlarged Hilbert space. For example, one may try to obtain the Bogolubov transformations between the different sectors.

Another direction is to tie up the loose - with regards to the boundary conditions - ends in obtaining the 1PI action and quantum-level field equations thereof.
Working out the explicit form of the reduced action for the {\em classical} action has its use, e.g., in making BHI and the BMS symmetry clearer. For the full dynamics of the bulk-boundary coupling, one will have to work out the 4D 1PI action and then reduce it to the 3D action.
For this, one should first obtain the offshell 4D 1PI action and then carry out the reduction of the physical states.
Additional boundary terms will be required to stay within the given boundary condition imposed at the classical level - whether it is a Dirichlet, Neumann or other boundary condition. Those and the surface terms generated by partial integrations must be kept track of to ensure consistency. The Dirichlet case has been worked out in section 2. The Neumann and other cases will be worth exploring. 
With the quantum-level boundary conditions checked, it will be interesting to extend to other cases the analysis carried out, e.g., in \cite{Nurmagambetov:2018het}, in which it was shown that an infalling observer will encounter a quantum-generated trans-Planckian energy near the horizon.

\newpage

\appendix

\renewcommand{\theequation}{A.\arabic{equation}}
\setcounter{equation}{0}

\section{Linear-order check of \rf{ptm} as a solution}

We have noted in section 2 that in a gravitational theory, consideration of both the active and passive transformations of the diffeomorphism symmetry leads to complementary pieces of the information about the boundary conditions and dynamics. In particular, the actively-transformed metric $g'_{\m\n}(x)$ implies a need for enlarging the Hilbert space.  
Here we explicitly check what may seem obvious: the metric given in \rf{ptm}, which we quote here for convenience, 
\be
g_{\m\n}'(x)= g_{0\m\n}(x)+h_{\m\n}(x)    \quad,\quad h_{\m\n} \equiv \nabla_\m \e_\n+\nabla_\n \e_\m  
\ee
satisfies the field equation since it is a gauge transformation of the solution $g_{0\m\n}(x)$. One may naively expect that $g_{\m\n}'(x)$ given above automatically (i.e., without any further condition on $\e^\m$) satisfies the field equation. As we will show now, the procedure actually reveals that $g_{\m\n}'(x)$ satisfies the metric field equation only when one gauge-fixes the fluctuation field $h_{\m\n}$. In other words, the parameter {$\e_\m$} must be restricted by the gauge-fixing conditions on $h_{\m\n}$.
We show this by employing a functional Taylor expansion. The reason we need such gauge-fixing must be that inversion of the kinetic operator - for which the gauge-fixing is required - is somehow built-in in the expansion through the functional Taylor derivative. 

Let us consider the $h_{\m\n}$-linear order of the Einstein equation:
\be
\int d^4 x'  h_{\a\b}     \fr{\d}{\d g_{\a\b}} \Big(R^{\m\n}-\fr12 g^{\m\n}R \Big)|_{g_{\a\b}=g_{0\a\b}} \nn\\
\ee
where {$|_{g_{\a\b}=g_{0\a\b}}$}, which will be suppressed from now on, denotes that $g_{0\a\b}$ is substituted {into} $g_{\a\b}$ after taking the derivative. After several straightforward steps one has
\bea
&&\hspace{-.3in}= \int   h_{\a\b} \Big( R^{\m\a\n\b}+g_{\r\s}\fr{\d}{\d g_{\a\b}}R^{\m\r\n\s} +\fr12 g^{\a\m}g^{\b\n} R+\fr12 g^{\m\n}R^{\a\b}-\fr12 g^{\m\n}g^{\r\s}\fr{\d R_{\r\s}}{\d g_{\a\b}} \Big). \nn\\
\eea
Since the background satisfies the Einstein equation it follows that $R=0=R^{\r\s}$ and one gets
\bea
&&\hspace{-.3in}= \int   h_{\a\b} \Big( R^{\m\a\n\b}+g_{\r\s}\fr{\d}{\d g_{\a\b}}R^{\m\r\n\s} -\fr12 g^{\m\n}g^{\r\s}\fr{\d R_{\r\s}}{\d g_{\a\b}} \Big).
\eea
Let us further evaluate the second and third terms above; by using
\bea
\d R^\r{}_{\s\m\n}= \nabla_\m \d\G^\r_{\n\s}-\nabla_\n \d\G^\r_{\m\s}\quad,\quad
\d R_{\m\n}=\nabla_\r \d\G^\r_{\m\n}-\nabla_\n \d\G^\r_{\r\m}  \la{rtv}
\eea
and
\bea
\d \G^\l_{\m\n}=\fr12 g^{\l\k}(\nabla_\m \d g_{\n\k}+\nabla_\n \d g_{\m\k}
- \nabla_\k \d g_{\m\n})
\eea
one can show that
\bea
 \int   h_{\a\b} g_{\r\s}\fr{\d}{\d g_{\a\b}}R^{\m\r\n\s} &=& \fr{1}2\Big( 2\nabla^\r\nabla^{(\n} h_\r^{\m)}-\nabla^\m\nabla^\n h_\g^\g-\N^2h^{\m\n} \Big) 
                                                                              -2R^{\m\a\n\b}-h^{\a\b}R^{(\m}_\a g_{\b}^{\n)} \nn\\
                                                                               &=& \fr{1}2\Big( 2\nabla^\r\nabla^{(\n} h_\r^{\m)}-\nabla^\m\nabla^\n h_\g^\g-\N^2h^{\m\n} \Big) 
                                                                              -2h_{\a\b}R^{\m\a\n\b} 
\eea
where in the second equality the Ricci tensor term has been omitted for the reason given above, and
\bea
 -\fr12 \int   h_{\a\b}g^{\m\n}g^{\r\s}\fr{\d R_{\r\s}}{\d g_{\a\b}} = - g^{\m\n}\Big( \N^p\N^q h_{pq}-\N^2 h_\g^\g \Big).
\eea
Let us now impose the following gauge conditions, which are basically the de Donder gauge with the additional tracelessness condition, 
\be
h_\g^\g=0\quad,\quad \N^\k h_{\k\m}=0.
\ee
With these, one gets
\be
\int   h_{\a\b} g_{\r\s}\fr{\d}{\d g_{\a\b}}R^{\m\r\n\s} 
= \fr{1}2\Big( 2\nabla^\r\nabla^{(\n} h_\r^{\m)}-\N^2h^{\m\n} \Big) 
-2h_{\a\b}R^{\m\a\n\b} \la{rtnz}
\ee
and
\be
-\fr12 \int   h_{\a\b}g^{\m\n}g^{\r\s}\fr{\d R_{\r\s}}{\d g_{\a\b}} = 0.
\ee
By substituting $h_{\m\n} \equiv \nabla_\m \e_\n+\nabla_\n \e_\m $ and using the following identities, 
\bea
\hspace{-.3in}(\N_\a \N_\b -\N_\b \N_\a)T^{\r_1\cdots \r_n}{}_{\l_1\cdots \l_m} &=& -\Sigma_{i=1}^n R_{\a\b\k}{}^{\r_i}T^{\r_1\cdots  \k\cdots \r_n}{}_{\l_1\cdots \l_m}\nn\\
 && +\Sigma_{j=1}^m R_{\a\b\l_j}{}^{\k}T^{\r_1\cdots  \r_n}{}_{\l_1\cdots \k \cdots \l_m} \nn\\
 \N_\a R_{\b\g\l}{}^\a&=& -\N_\b R_{\g\l}+\N_\g R_{\b\l}, 
\eea
the first two terms in \rf{rtnz} can be evaluated further.
First note that  the gauge conditions translate into
\be
\N^\k \e_\k=0\quad,\quad \N^2 \e_\n=0.
\ee
For instance, {one of the terms to be computed is}
\bea
\N^\r \N^\n \N_\r \e^m &=& \Big([\N^\r, \N^\n] +\N^\n \N^\r  \Big) \N_\r \e^\m \nn\\
                                 &=& -R^{\k_1\n\k_2\m}\N_{\k_1} \e_{\k_2}
\eea
where the second equality was obtained after using $R^{\m\n}=0$ and $\N^2 \e^\m=0$.
By evaluating the other terms in \rf{rtnz} and using the identities given above, one can show that  \rf{rtnz} vanishes, completing the proof.

\end{document}